\documentclass{article}
\usepackage{spconf,amsmath,graphicx}
\usepackage{amsfonts,amssymb} 

\title{Fine-grained Disentangled Representation Learning for Multimodal Emotion Recognition}

\name{Haoqin Sun$^1$, Shiwan Zhao\sthanks{Independent researcher.}, Xuechen Wang$^1$, Wenjia Zeng$^2$, Yong Chen$^2$, Yong Qin$^{1}$\sthanks{Corresponding author. This work was supported in part by NSF China (Grant No. 62271270)}}
\address{$^1$  Nankai University, Tianjin, China\\$^2$  Lingxi (Beijing) Technology Co., Ltd.}

\begin{document}
\ninept
\maketitle
\begin{abstract}
Multimodal emotion recognition (MMER) is an active research field that aims to accurately recognize human emotions by fusing multiple perceptual modalities. However, inherent heterogeneity across modalities introduces distribution gaps and information redundancy, posing significant challenges for MMER. In this paper, we propose a novel fine-grained disentangled representation learning (FDRL) framework to address these challenges. Specifically, we design modality-shared and modality-private encoders to project each modality into modality-shared and modality-private subspaces, respectively. In the shared subspace, we introduce a fine-grained alignment component to learn modality-shared representations, thus capturing modal consistency. Subsequently, we tailor a fine-grained disparity component to constrain the private subspaces, thereby learning modality-private representations and enhancing their diversity. Lastly, we introduce a fine-grained predictor component to ensure that the labels of the output representations from the encoders remain unchanged. Experimental results on the IEMOCAP dataset show that FDRL outperforms the state-of-the-art methods, achieving 78.34\% and 79.44\% on WAR and UAR, respectively.
\end{abstract}
\section{Introduction}
\label{sec:intro}

Multimodal emotion recognition (MMER) is an important subfield of affective computing that aims to utilize information from speech and other perceptual modalities, such as textual expressions and body language, to recognize and comprehend human emotional states. In recent years, MMER has found notable applications in various domains, including human customer service \cite{lee2005toward}, robotics \cite{michaud2000artificial}, and car driving \cite{schuller2004speech}. However, the inherent heterogeneity among modalities results in inconsistency and imbalance of information, thereby increasing the difficulty of multimodal representation learning and fusion.

The current mainstream research primarily focuses on two directions. First, a considerable amount of effort has been dedicated to designing intricate fusion strategies for obtaining effective multimodal representations, which in turn facilitate a more comprehensive understanding of emotions and enhance the performance of multimodal technologies \cite{wang2019words,zhang2022tailor}.
For instance, Liu et al. \cite{liu2022multi} proposed a multi-scale fusion framework that combines feature level and decision level fusion for achieving multimodal interaction between speech and text. However, their approach directly concatenates the representations of the two modalities, thereby ignoring the heterogeneity across them.
Second, researchers have been committed to exploring the correlations between different modal information to obtain more robust modal semantics and, consequently, improve the accuracy of emotion recognition tasks \cite{lv2021progressive,tsai2019multimodal,ioannides23_interspeech}. For example, Hazarika et al. \cite{hazarika2020misa} utilized difference loss to promote the orthogonality of features, thereby increasing the multimodal complementary information. Nonetheless, this might not be an ideal solution as the aforementioned difference loss alone is insufficient for obtaining reliable representations and may lead to trivial solutions without appropriate constraints. 
More recently, Yang et al. \cite{yang2022disentangled} proposed a method for disentangled representation learning, which extracts both common and private feature representations for each modality. While this approach holds promise, it primarily focuses on global modality alignment and disparity, neglecting fine-grained feature disentanglement. This oversight limits the model's ability to capture the nuances of each modality, resulting in only suboptimal performance.

To address the limitations of existing approaches and facilitate fine-grained feature disentanglement, we introduce a novel fine-grained disentangled representation learning (FDRL) framework, designed to address the heterogeneity of modalities. This process consists of two main steps. 
First, we learn modality-shared representations where a shared encoder projects each modality into a modality-shared subspace. Here, a fine-grained alignment component is employed to balance global and local distribution alignment, capturing modality consistency.
Second, we learn modality-private representations where private encoders map each modality, separately, into a modality-private subspace. To ensure diversity, a fine-grained disparity component constrains the private latent spaces. 
Moreover, we leverage a fine-grained predictor component to ensure that the label associated with the output representations remains unchanged. Finally, the cross-modal fusion (CMF) module is utilized to efficiently fuse the different representations for MMER. Experimental results on the IEMOCAP dataset demonstrate that our method surpasses state-of-the-art methods, achieving 78.34\% and 79.44\% on WAR and UAR, respectively.

\begin{figure*}[t]
  \centering
  \includegraphics[width=6.0in]{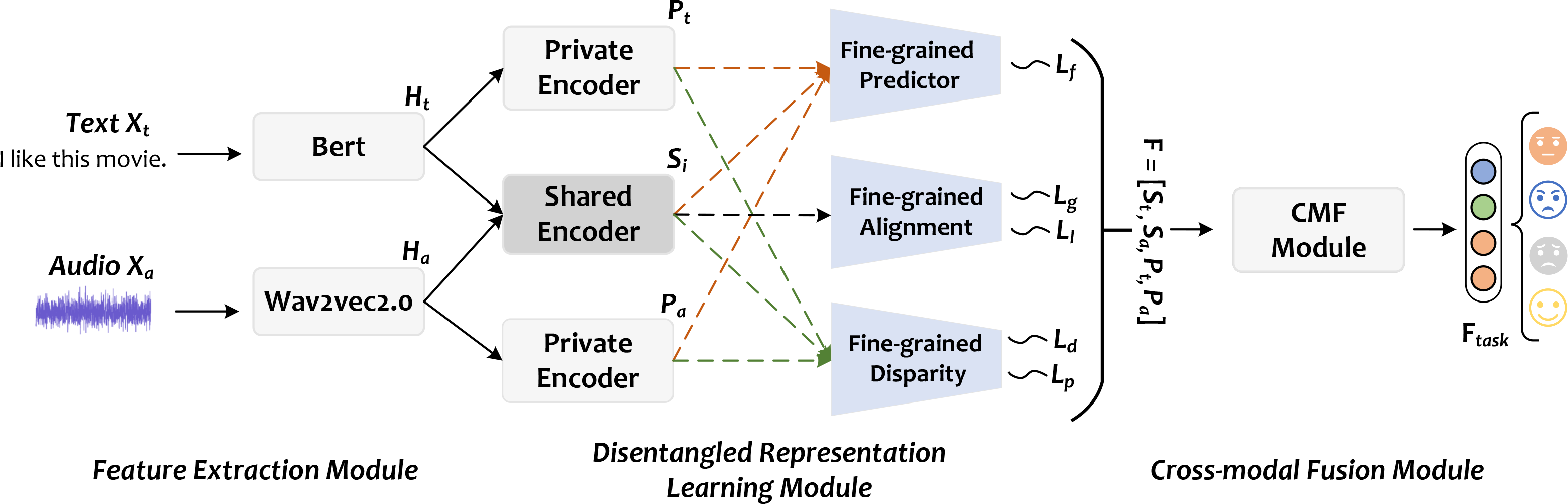}
  \caption{The overall structure of the proposed framework.}
  \label{fig:framework}
\end{figure*}

Our main contributions are summarized as follows:

We propose the FDRL framework, a fine-grained disentangled representation learning method for MMER. FDRL focuses on achieving fine-grained feature disentanglement to alleviate the modal heterogeneity gap.

We introduce fine-grained alignment, disparity, and predictor components to learn refined disentangled features. These components are employed to adaptively align feature distributions, reduce information redundancy, increase feature diversity, and guide the learning of representations.

Experimental results on the benchmark dataset, IEMOCAP, demonstrate the effectiveness of our proposed method. 

\section{The proposed method}
\label{sec:format}

As shown in Fig. \ref{fig:framework}, our proposed method consists of three modules: the feature extraction module, the disentangled representation learning module, and the cross-modal fusion (CMF) module. Notably, the disentangled representation learning module contains three key components: the fine-grained alignment component, the fine-grained disparity component, and the fine-grained predictor component.

\subsection{Feature Extraction Module}\label{sec:Extraction}

The sequence representations of the speech and text modalities are denoted as $X_a \in R^{L_a\times d_a}$ and $X_t \in R^{L_t\times d_t}$, respectively, where $L$ represents the sequence length and $d$ represents the embedding dimension. To obtain the high-level representation of each modality $H_i, i \in \{a,t\}$, we utilize Wav2vec2 \cite{baevski2020wav2vec} and Bert \cite{devlin2018bert} encoders to enrich the modal information:
\begin{align*}
H_{a}=\text {Wav2vec2}\left(X_{a} ; \theta_{a}^{\text {wav}}\right) \in R^{L_{a} \times d_{a}},\tag{1}\\
H_{t}=\text {Bert}\left(X_{t} ; \theta_{t}^{\text {bert}}\right) \in R^{L_{t} \times d_{t}},\tag{2}
\end{align*}
where $\theta_{a}^{\text {wav}}$ and $\theta_{t}^{\text {bert}}$ represent the trainable parameters. 

\subsection{Disentangled Representation Learning Module}\label{sec:Learning}

\textbf{Shared and Private Encoders:} The heterogeneity of modalities leads to information redundancy and distribution gaps in the high-level representations. To address these issues, we propose using shared and private encoders to capture the commonality and complementarity of heterogeneous modalities. Specifically, the encoders project the features into shared and private feature subspaces. Both the shared encoder $\mathbb E_{s} (\cdot; \theta_s)$ and the private encoder $\mathbb E_{i} (\cdot; \theta_i), i \in \{a,t\}$ are implemented using two fully connected layers with the ReLU activation function. With the encoders, the shared and private representations can be computed as follows:
\begin{align*}
S_{i}=\mathbb E_{s}\left(H_{i} ; \theta_{s}\right),P_{i}=\mathbb E_{i}\left(H_{i} ; \theta_{i}\right),\tag{3}
\end{align*}
where $S_i$ and $P_i \in R^{d}$. $\theta_{s}$ and $\theta_{i}$ are trainable parameters of the encoders. The parameters $\theta_{s}$ are shared among all modalities, whereas $\theta_{i}$ is learned separately for each modality.

\begin{figure}[t]
  \centering
  \includegraphics[width=2.5in]{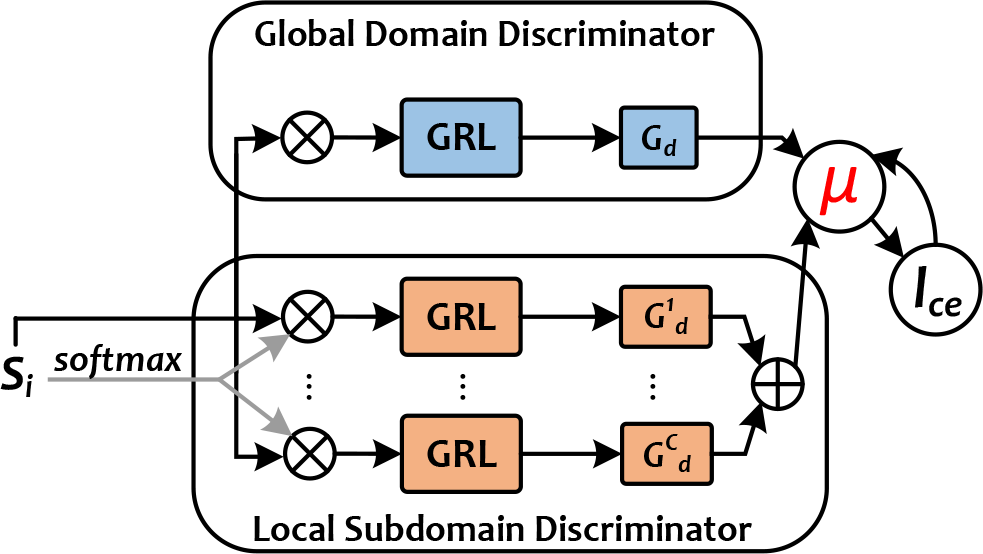}
  \caption{Detail of the fine-grained alignment component. GRL represents the gradient reversal layer and $\mu$ the dynamic factor.}
  \label{fig:framework1}
\end{figure}

\textbf{Fine-grained Alignment:} Inspired by the dynamic adversarial adaptation network (DAAN) \cite{yu2019transfer}, this component is designed to align the shared representations $S_i$. It comprises three elements: the global domain discriminator, the local subdomain discriminator, and the dynamic factor, as depicted in Fig. \ref{fig:framework1}.

(i) Global domain discriminator: The blue part in Fig. \ref{fig:framework1} represents the global domain discriminator $G_{d}$, which is utilized to align the global distribution of the speech and text modalities. The structure of the global domain discriminator is inspired by the domain adversarial neural network (DANN) \cite{ganin2016domain}, comprising two layers of perceptrons activated by the ReLU function. The loss of the global domain discriminator is calculated as follows:
\begin{align*}
L_{g}=\sum_{i \in a \cup t}l_{ce}\left(G_{d}\left(S_i\right), y_{m}\right),\tag{4}
\end{align*}
where $l_{ce}$ denotes the cross-entropy loss (domain discriminator loss), and $y_{m}$ represents the ground-truth modality label.

(ii) Local subdomain discriminator: The orange part in Fig. \ref{fig:framework1} represents the local subdomain discriminator $G_d^c$, which is used to align the local distribution of the speech and text modalities. The local subdomain discriminator consists of two-layer perceptrons with the activation function of ReLU. Specifically, local subdomain discriminators are associated with categories. Each category corresponds to a domain discriminator $G_d^c$. The output of softmax indicates the degree of attention given to $S_i$ by the domain discriminator. The loss of the local subdomain discriminator is computed as follows:
\begin{align*}
L_{l}=\sum_{c=1}^C\sum_{i \in a \cup t}l_{ce}^c\left(y_i^cG_{d}^c\left(S_i\right), y_{m}\right),\tag{5}
\end{align*}
where $l_{ce}^c$ denotes the cross-entropy loss for category $c$, and $y_i^c$ represents the predicted probability distribution of category $c$ for $S_i$.

(iii) Dynamic factor: The core of the fine-grained alignment component is the dynamic factor $\mu$, which dynamically estimates the relative importance of global and local distributions. The global $\mathcal A$-distance of the global domain discriminator is calculated as:
\begin{align*}
d_{\mathcal{A}, g}\left(\mathcal{M}_{a}, \mathcal{M}_{t}\right)=2\left(1-2\left(L_{g}\right)\right),\tag{6}
\end{align*}
where $\mathcal{M}{a}$ and $\mathcal{M}{t}$ denote the speech and text modalities.

The local $\mathcal A$-distance of the local domain discriminator is:
\begin{align*}
d_{\mathcal{A}, l}\left(\mathcal{M}_{a}^c, \mathcal{M}_{t}^c\right)=2\left(1-2\left(L_{l}^c\right)\right),\tag{7}
\end{align*}
where $\mathcal{M}_{a}^c$ and $\mathcal{M}_{t}^c$ are samples from category $c$ of the speech and text modalities, and $L_{l}^c$ is the local subdomain discriminator loss for category $c$. Finally,the dynamic factor $\mu$ is computed as:
\begin{align*}
\mu=\frac{d_{\mathcal{A}, g}\left(\mathcal{M}_{a}, \mathcal{M}_{t}\right)}{d_{\mathcal{A}, g}\left(\mathcal{M}_{a}, \mathcal{M}_{t}\right)+\frac{1}{C} \sum_{c=1}^{C} d_{\mathcal{A}, l}\left(\mathcal{M}_{a}^{c}, \mathcal{M}_{t}^{c}\right)}.\tag{8}
\end{align*}

\textbf{Fine-grained Disparity:} This component consists of two parts: global modality discriminator and local orthogonal constraint.

(i) Global modality discriminator: To generate refined private representations and enhance feature diversity, we encourage the global modality discriminator $\ell_{GD}(\cdot;\theta_{GD})$ to distinguish the source of modality:
\begin{align*}
L_{p}=\frac{1}{2}\sum_{i \in a \cup t}l_{ce}\left(\ell_{GD}(P_i;\theta_{GD}), y_m\right),\tag{9}
\end{align*}
where $\theta_{GD}$ are trainable parameters. The global modality discriminator comprises two-layer perceptrons with a ReLU activation function.

(ii) Local orthogonal constraint: We apply a disparity loss to encourage the orthogonalization of shared and private features, which reduces information redundancy:
\begin{align*}
L_{d}=\left\|S_{a}^{\top} P_{a}\right\|_{F}^{2} + \left\|S_{t}^{\top} P_{t}\right\|_{F}^{2},\tag{10}
\end{align*}
where $\|\|_{F}^{2}$ denotes the squared Frobenius norm. The local orthogonal constraint prevents the generation of trivial solutions during the process of encouraging feature orthogonalization.

\textbf{Fine-grained Predictor:}
Our shared and private encoders only need to learn how to move the feature vectors in the feature space. It is worth noting that the mentioned restriction strategy alone is not sufficient to obtain reliable encoders, since it may randomly manipulate the representation without proper constraints. To this end, we require the fine-grained predictor component $\ell_{FP}(\cdot;\theta_{FP})$ to restrict that the output representation of the encoder should not change the associated label. 
\begin{align*}
L_{f}=\frac{1}{3}\sum_{j \in S\cup P_a \cup P_t}l_{ce}\left(\ell_{FP}(j;\theta_{FP}), y_e\right), \tag{11}
\end{align*}
where $S$ represents the sum of $S_a$ and $S_t$. $y_e$ represents the ground-truth emotion label. $\theta_{FP}$ is trainable parameters. The fine-grained predictor component is a one-layer perceptron.

\subsection{Cross-Modal Fusion Module}
After obtaining these modal representations, we first stack them into a matrix $\mathbf{F} = [S_a,S_t,P_a,P_t] \in R^{4 \times d}$. Then, we use the multi-head self-attention mechanism \cite{vaswani2017attention} to obtain the final cross-modal representation.
\begin{align*}
\text{Attention}(Q, K, V)=\operatorname{softmax}\left(\frac{\mathbf{Q K}^{T}}{\sqrt{d_{h}}}\right) \mathbf{V}, \tag{12}\\
head_{i}=\text { Attention }\left(\mathbf{Q} W_{i}^{q}, \mathbf{K} W_{i}^{k}, \mathbf{V} W_{i}^{v}\right), \tag{13}
\end{align*}
where $\mathbf{Q}=\mathbf{K}=\mathbf{V}=\mathbf{F}$. $W$ is trainable parameters. Finally, we concatenate each head to obtain the representation $\mathbf{F}_{task}$ for the downstream task. The task loss and final loss are calculated as follows:
\begin{align*}
L_{task}=l_{ce}\left(\mathbf{F}_{task}, y_e\right), \tag{14}\\
L_{total}= L_{task} + \alpha[(1-\mu) L_{g}+ \mu L_{l}]\\+ \beta( L_{p}+L_{d}) + \gamma L_{f}, \tag{15}
\end{align*}
where $\alpha, \beta$, and $\gamma$ represent the trade-off parameters.

\section{Experiments}
\label{sec:experiment}

\subsection{Emotion dataset}
The Interactive Emotional Dyadic Motion Capture (IEMOCAP) dataset \cite{busso2008iemocap} is a widely used multimodal database for research in emotion recognition and analysis. The dataset is designed to facilitate research in automatic emotion recognition, sentiment analysis, and other related fields using multimodal data. The dataset is collected from ten actors, who participated in improvised sessions to simulate natural interactions. Our experiment considers four emotions: angry, happy, sad and neutral, where excitement class is merged into the happy class.

\subsection{Experimental setup}
We employ 5-fold cross-validation to evaluate our proposed framework, with four sessions for training and one session for testing. The batch size is set to 8 and the maximum training epoch is set to 100. We choose the AdamW optimizer with an initial learning rate of $10^{-5}$. The trade-off parameters $\alpha=\gamma=1$ and  $\beta=0.5$. We use the weighted average recall (WAR) and unweighted average recall (UAR) as metrics to evaluate the performance of our proposed method.

\subsection{Comparison with State-of-the-Art Methods}
To evaluate the effectiveness of our proposed method, we compare it with various state-of-the-art (SOTA) methods, and the experimental results are listed in Table 1. We observe that our method achieves the best performance in both WAR and UAR metrics, significantly outperforming the previous SOTA methods. This superior performance is mainly attributed to our method's exceptional ability to focus on learning the consistency and complementarity of heterogeneous modalities. These results provide strong evidence that our proposed fine-grained disentangled representation learning method can deliver promising performance for MMER.
\begin{table}[t]
\caption{Classification performance of various state-of-the-art approaches on the IEMOCAP dataset.}
\center
\begin{tabular}{lcc}
\hline
Methods                    & WAR(\%) & UAR(\%) \\\hline
BERT + FBK \cite{morais2022speech}           & 70.56  & 71.46  \\
SMCN\cite{hou2022multi}     & 73.50  & 73.00  \\
MCNN+BLSTM \cite{liu2022multi}             & 74.98  & 75.05  \\
BERT + Wav2vec2 \cite{zhao2022multi}   & -      & 76.31  \\
Multi-granularity model \cite{fan2023mgat} & 74.80  & 75.50  \\ 
SDTF-NET \cite{liu2023sdtf} & 77.20  & 76.70  \\
MISA \cite{hazarika2020misa} & 76.18  & 75.73  \\
Co-att + BAM+knowledge\cite{zhao2023knowledge}        & 75.50  & 77.00  \\\hline
\textbf{Ours}              & \textbf{78.34}  & \textbf{79.44}  \\ \hline
\end{tabular}
\end{table}

\subsection{Ablation Studies}

We conduct ablation studies on the IEMOCAP dataset to assess the importance of different components in FDRL. Table 2 shows the results with the following observations. 

\textbf{Importance of Regularization:} 
Without the fine-grained alignment and disparity components (S1), our framework only learns the representations of different modalities without feature disentanglement. This results in poor performance as it fails to overcome the heterogeneity of the modalities and the diversity of modal information. Moreover, we observe that constraining the encoders with the fine-grained predictor component (S6) also enhances the framework's performance.

\textbf{Importance of Global/Local discriminator:} 
We evaluate the impact of global and local discriminators in the fine-grained alignment component by comparing the performance of S0 with S2 and S0 with S3. The FDRL framework improves by 0.95\% on WAR and 1.18\% on UAR compared to S2, and by 0.83\% on WAR and 1.00\% on UAR compared to S3. This indicates that merely aligning the global distributions of different modalities, although beneficial, is insufficient for learning better representations. Furthermore, aligning the local distributions of different modalities is crucial for learning more refined representations.

\textbf{Importance of Different Representations:} 
By keeping the entire framework but using only partial representations in the CMF module (S4/S5), we observe that our proposed method outperform S4 and S5. This underscores the importance and necessity of learning both shared and private representations. Notably, private representations are found to be more expressive than shared representations.

\subsection{Visualization}
\begin{figure}[t]
  \centering
  \includegraphics[width=0.48\textwidth]{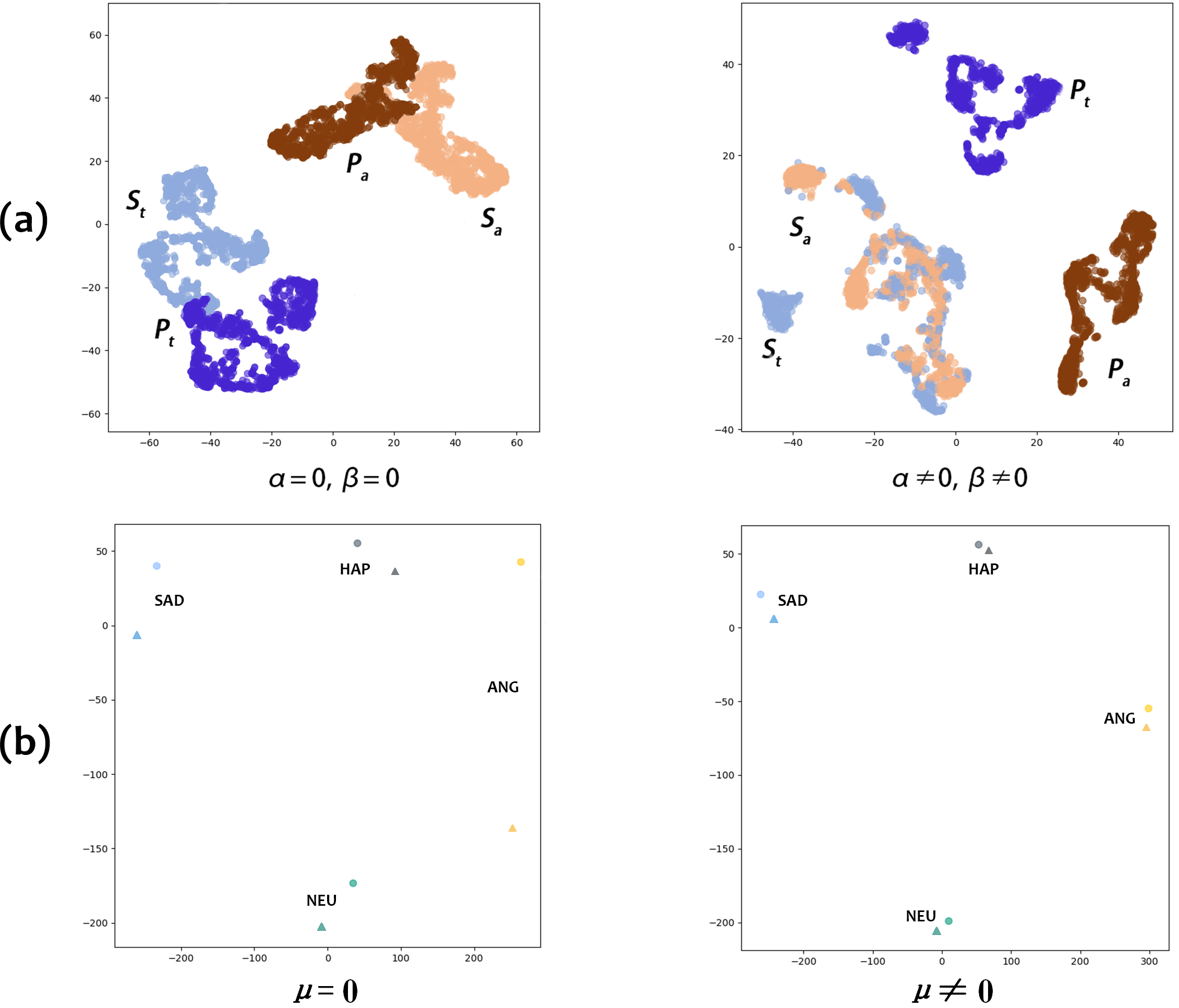}
\caption{(a) Visualization of the shared and private representations in the testing set on the IEMOCAP corpus. The fine-grained alignment and disparity components pull the shared representations closer while disentangling shared and private representations. (b) Visualization of the centroids of the shared representations $S_i$ for each category within the text and speech modalities.}
\label{fig:TSNE}
\end{figure}

In Fig. \ref{fig:TSNE} (a), we use t-SNE \cite{van2008visualizing} to visualize the distributions of shared and private representations with and without the fine-grained alignment and disparity components. When $\alpha = 0$ and $\beta = 0$, the distributions of $S_a$ and $S_t$ do not overlap, while the distributions of $S_i$ and $P_i$ sometimes overlap, indicating that disentangled representations are not learned. In contrast, when $\alpha \neq 0$ and $\beta \neq 0$, the distributions of $S_i$ begin to mix and overlap, while the distribution of $P_i$ becomes progressively more distinct. This suggests that our method captures the commonality and complementarity of heterogeneous modalities, with the fine-grained alignment component minimizing the distribution between modalities and the fine-grained disparity component maximizing the feature diversity of modalities.

In Fig. \ref{fig:TSNE} (b), we visualize the centroids of the shared representations $S_i$ for each category of text and speech modalities to further demonstrate the effectiveness of aligning local subdomain distributions (associated with categories) in the fine-grained alignment component. When $\mu = 0$, the feature centroids of the same emotion from different modalities are distant from each other, indicating that shared features of local subdomains are not learned. When $\mu \ne 0$, the feature centroids of the same emotion gradually come closer together. This suggests that the distribution of local subdomains is brought closer by the fine-grained alignment component, which helps improve the performance of the framework.

\begin{table}[t]
\caption{Results of ablation studies on IEMOCAP dataset (WAR and UAR in \%).}
\vspace{-3mm}
\center
\begin{tabular}{clll}
\hline
System&\multicolumn{1}{c}{Model}                                                  & WAR  & UAR  \\ \hline
\textbf{S0}&\multicolumn{1}{c}{\textbf{FDRL}}                    & \textbf{78.34}  & \textbf{79.44}   \\ \hline
S1&\multicolumn{1}{l}{w/o ($L_g$ + $L_l$) + ($L_p$+ $L_{d}$)}                   &76.47     & 77.36    \\
S2&\multicolumn{1}{l}{w/o Global discriminator}            &77.39     & 78.26               \\
S3&\multicolumn{1}{l}{w/o Local discriminator}                  &77.51     & 78.44        \\
S4&\multicolumn{1}{l}{w/o fine-grained alignment component}                  &77.15     & 77.43        \\
S5&\multicolumn{1}{l}{w/o fine-grained disparity component}            &76.69     & 78.01  \\
S6&\multicolumn{1}{l}{w/o fine-grained predictor component}            &77.71     & 78.67  \\
\hline
\end{tabular}
\end{table}

\section{Conclusions}
In this paper, we propose a fine-grained disentangled representation learning framework to deal with the heterogeneity of modalities by disentangling the features of each modality. Our framework learns shared representations of heterogeneous modalities by capturing the consistency of global and local distributions of modalities through a fine-grained alignment component. Subsequently, we utilize the fine-grained disparity component to enhance the quality of private representations, thus reducing information redundancy. The fine-grained predictor is employed to ensure the encoder does not alter the associated label. Finally, the CMF module is used to generate refined cross-modal representations. The effectiveness of our proposed method is demonstrated through a series of comparative experiments and ablation studies on the IEMOCAP dataset. 

\bibliographystyle{IEEEbib}
\bibliography{strings,refs}

\end{document}